\documentstyle[bo99,epsfig]{article}

\title{High Resolution Observations of X-Ray Absorbers/Emitters}
\author{Fabrizio Nicastro$^{1,2,3}$, Martin Elvis$^{1}$, 
Fabrizio Fiore$^2$, Giorgio Matt$^4$ and Sandra Savaglio$^2$}
\affil {$^1$ Harvard-Smithsonian Center for Astrophysics, 
60 Garden st. Cambridge MA. 02138 USA. $^2$ Osservatorio 
Astronomico di Roma, via Osservatorio, Monteporzio-Catone (RM), 
I00040 Italy. $^3$ Istituto di Astrofisica Spaziale - CNR, 
Via del Fosso del Cavaliere, Roma, I-00133 Italy. $^4$ Department 
of Physics, Universit\'a degli Studi Roma Tre Via della Vasca 
Navale 84, Roma I-00146, Italy} 

\begin{document}

\maketitle

\begin{abstract}
We present photoionization and collisional ionization models, 
and their application to three important fields: (a) the Warm 
Absorbers/Emitters in type 1 AGN, (b) the Warm Reflectors 
in Type 2 AGN, and (c) X--ray absorption of background quasars 
by intergalactic gas. 
A number of cases are investigated, and the dependences of the main 
parameters explored. 
\end{abstract}

\section{Introduction}
Two, apparently distinct, main components are resolved in 
low resolution X-rays spectra of AGN: (a) the so called ``warm absorber'' 
(and possibly emitter), found in half of the Seyfert 1 galaxies observed 
by ASCA (Reynolds, 1997; George et al., 1998), and (b) the ionized reflector, 
seen in a number of Seyfert 2 galaxies observed by ASCA and BeppoSAX (i.e. 
Turner et al., 1997; Comastri et al., 1998; Guainazzi et al., 1999). 

Ionized matter, potentially absorbing background quasar radiation, 
is also expected to be present (maybe under the form of filaments) 
in the intergalactic space, and its presence has relevant cosmological 
consequences (Hellsten, Gnedin \& Miralda-Escud\'e, 1998). 
This matter is expected to produce strong resonant absorption lines in the 
X-ray band (i.e. ``X-ray Forest''), which can be used as powerful 
diagnostics of the ionization state, and the temperature of this diffuse 
gas. 

\section{The Models}
Our models for photoionized and collisionally ionized gas (for a 
detailed presentation see Nicastro, Fiore \& Matt, 1999: NFM99; 
Nicastro, Fiore, Matt \& Elvis, in preparation: NFME99), include all 
the strongest (oscillator strength $> 0.1$) absorption lines as well 
as permitted, intercombination and forbidden emission lines in the 50 
eV to 10 keV band. 
The ionization structure of the gas is computed by using CLOUDY 
(vs. 90.04, Ferland, 1996). 
Resonant absorption is included as in NFM99, while the emission 
contribution is that predicted by CLOUDY. The lines profile (both 
in emission and absorption) is the correct voigt profile (NFM99). 
The geometrical configuration of the absorbing/emitting clouds is 
properly accounted for by weighting the relative (absorption versus 
emission) intensity with the covering factor $f_c$ as seen by the 
central source (for details see NFME99). 

\section{Warm Absorbers/Emitters in Type 1 AGN}
Figure 1a (left panel) shows two spectra reprocessed by: (1) photoionized 
outflowing and turbulent gas, with $v_{out} = 1,000$ km s$^{-1}$, 
$v_{turb} = 500$ km s$^{-1}$, $f_c = 0.5$, log~N$_H$ = 22 (in cm$^{-2}$), 
log~n$_H$ = 10 (in cm$^{-3}$), log~U = 0.5, and equilibrium temperature of 
$T = 4.5 \times 10^4$ K (upper panel); (2) gas with the same 
dynamical/geometrical parameters and densities, but with log~U = -0.2 and 
$T = 3.2 \times 10^6$ K (lower panel). 
In the latter case the gas is not in photoionization equilibrium: the 
temperature is kept higher by an external source of heating (as suggested 
in the case of the ``truly-warm'' absorber of NGC~5548: Nicastro et al., 
1999). The value of U in the non-equilibrium case (lower panel) has 
been chosen to give OVII-OVIII relative abundances similar to the pure 
photoionization case. 
We note that the emissivity of the gas is strongly enhanced in the 
non-equilibrium case, and the OVIIK$\alpha$ triplet is now clearly 
visible.
The right panel (Fig. 1b) shows 100 ks Chandra-MEG simulations 
of the models of Fig. 1a. The 2-10 keV source flux is of 1 mCrab 
$ = 2 \times 10^{-11}$ erg s$^{-1}$ cm$^{-2}$. 
The Chandra-MEG, clearly resolves 
most of the 0.5-1 keV absorption and emission lines by highly ionized 
Oxygen and Neon, and allows one to measure their relative intensity and 
width. The OVIIK$\alpha$ triplet is clearly resolved in the spectrum 
reprocessed by gas with $T = 3.2 \times 10^6$ K. 

\vspace{-2truecm}
%
\begin{figure}
\centerline{\psfig{file=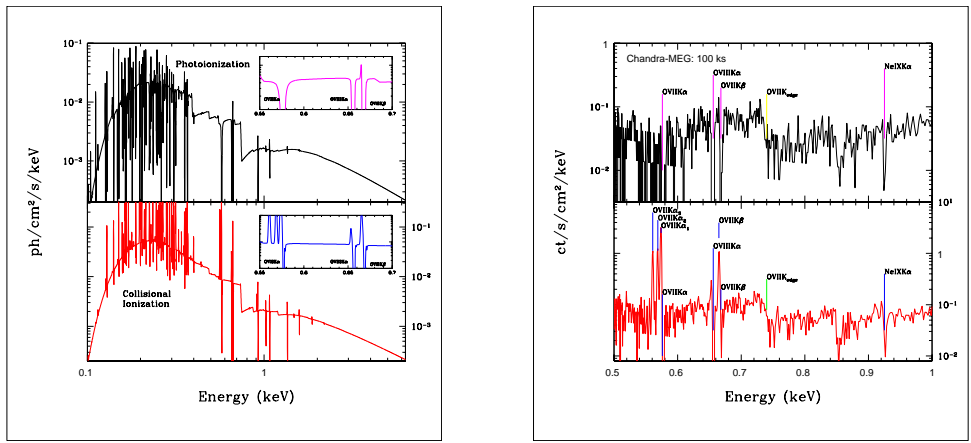}}
\caption[]{Left panel: pure photoionization (upper panel) and mix 
photoionization/collisional ionization (lower panel) models for Warm 
Absorber/Emitters in type 1 AGN. Right panel: 100 ks 
Chandra-MEG simulations of the models in the left panel.}
\end{figure}
%
%
\begin{figure}
\centerline{\psfig{file=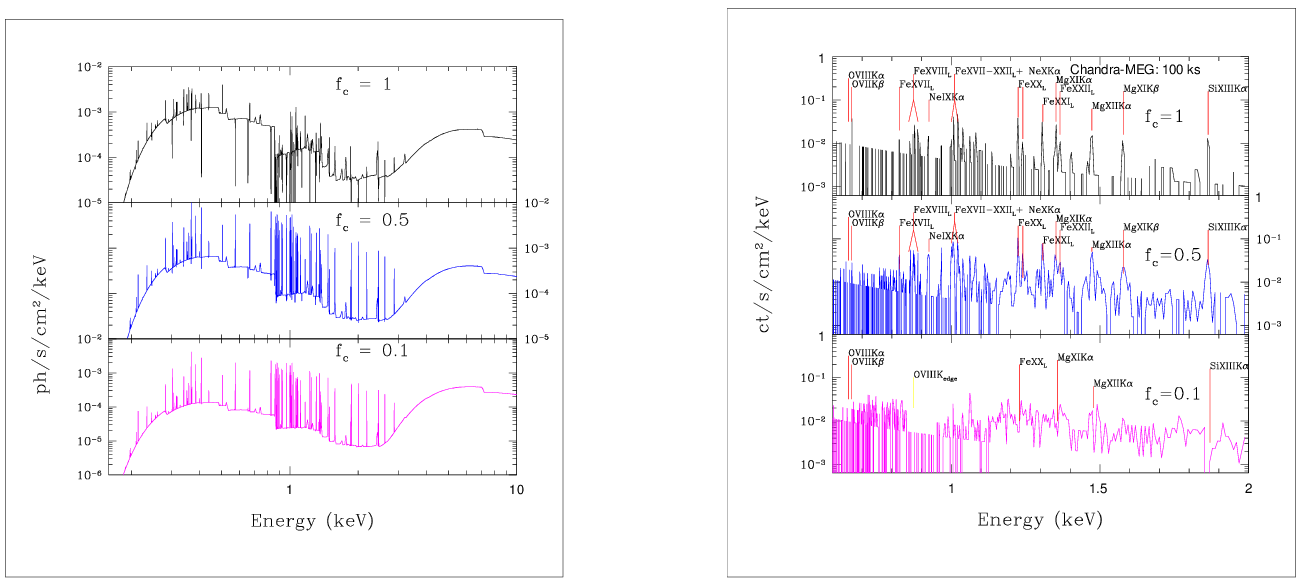}}
\caption[]{Left panel: Photoionization models for ``warm-reflectors'' 
in Seyfert 2s. Right panel: 100 ks 
Chandra-MEG simulations of the models in the left panel.}
\end{figure}
%

\vspace{2truecm}
\section{Warm Reflectors in type 2 AGN}
Figure 2a (left panel) shows the same photoionzation model as in 
the upper panel of Figure 1a, except the direct nuclear continuum is 
now obscured by a column of neutral gas of N$_H^{Cold} = 3 \times 
10^{23}$ cm$^{-2}$ (Seyfert 2-like). Only the {\em Warm-Reflector} is 
visible in this case. The three cases correspond to three different 
values of the covering factor of the warm medium as seen by the 
central (obscured) source. 
Figure 2b (right panel) shows 100 ks Chandra-MEG simulations 
of the models in Fig. 2a. The 2-10 keV source flux is of 1 mCrab. 
The Chandra-MEG resolves most of the predicted 0.5-2 keV emission 
lines by O, Ne, Mg, Si and Fe highly ionized.

\section{Absorption Line Diagnostics of Hot Intergalactic Plasma}
Figure 3a (left panel) shows two spectra of a bright background 
quasar transmitted by diffuse hot gas with log~N$_H = 21$ (in cm$^{-2}$), 
log~n$_H$ = -3 (in cm$^{-3}$), and two different temperatures: log~T = 6.5 
(upper panel), and log~T = 7 (lower panel). 
Highly ionized oxygen produces the strongest absorption features 
from the $10^{6.5}$ K plasma, while L absorption lines by FeXVII-XVIII 
are imprinted on spectra transmitted by the hotter gas.
Figure 3b (right panel) shows 100 ks Chandra-MEG simulations of the 
models of Figure 3a. 
The 2-10 keV flux of the background quasar is of 1 mCrab. 
The Chandra-MEG resolves most of the predicted 0.5-1 keV absorption 
lines, and clearly allows one to distinguish the two considered cases: 
the intensities of OVII-VIII K$\alpha,\beta$ and NeIX K$\alpha$ 
absorption lines are a very powerful diagnostics.  
%
\begin{figure}
\centerline{\psfig{file=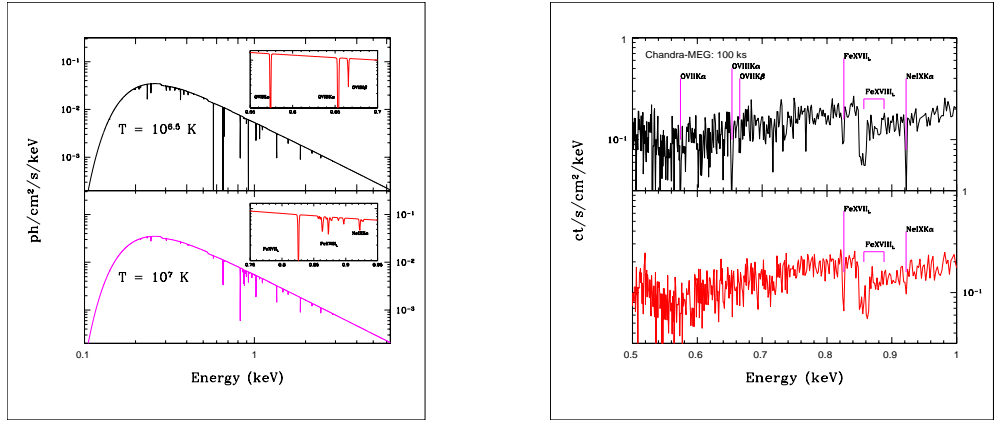}}
\caption[]{Left panel: models for absorption of a bright 
background quasar from hot intracluster gas. 
Right panel: 100 ks Chandra-MEG simulations of the models in the 
left panel.}
\end{figure}
%

\section{Conclusions}
We have presented our photoionization and collisional ionization 
models, and briefly discussed three different applications 
to as many important astronomical fields. Simulations with the 
Chandra-MEG of each of these cases have been shown. 

\begin{acknowledgements}
F.N. thanks H. Netzer and I.M. George for the useful 
discussions during the meeting. This work has been 
partly supported by the NASA grant NAG5-2476. 
\end{acknowledgements}

\end{document}